\documentclass[10pt]{article}
\usepackage[letterpaper]{geometry}
\usepackage{hicss}
\usepackage{times}
\usepackage[none]{hyphenat}
\usepackage{url}
\usepackage{latexsym}
\usepackage{minted}
\usepackage{indentfirst}
\usepackage{graphicx}
\graphicspath{{images/}}
\usepackage[
    style=apa,
  ]{biblatex}
\usepackage{booktabs}
\usepackage{array}
\usepackage{tabularx}
\usepackage{amsmath}
\usepackage{mdframed}
\usepackage{xcolor}

\newcommand{\code}[1]{\texttt{\small #1}}

\newmdenv[
 linewidth=0.5pt,
  linecolor=black!55,
  backgroundcolor=black!4,
  innertopmargin=3pt, innerbottommargin=3pt,
  innerleftmargin=5pt, innerrightmargin=5pt,
  skipabove=4pt, skipbelow=4pt,
  fontcolor=black
]{operatorprompt}
\newmdenv[
  linewidth=0.5pt,
  linecolor=black!55,
  backgroundcolor=white,
  innertopmargin=3pt, innerbottommargin=3pt,
  innerleftmargin=5pt, innerrightmargin=5pt,
  skipabove=4pt, skipbelow=4pt
]{agentresponse}

\title{CONDUCTOR: An LLM-Orchestrated Digital Twin for Uncertainty-Aware Distribution Grid Operations}
\author{Antonio Alc\'antara, Aysegül Kahraman, Anosh Arshad Sundhu, Spyros Chatzivasileiadis \\
  Department of Wind and Energy Systems, Technical University of Denmark (DTU) \\
  {\underline{\{anmata, ayska, anosu, spchatz\}@dtu.dk}} }

\date{}

\begin{document}
\maketitle
\begin{abstract}

%up to 150 words in length. 
Large language models (LLMs) are proposed as natural-language interfaces to power system analysis, yet existing frameworks are validated almost exclusively on synthetic benchmarks and support only deterministic studies. We present CONDUCTOR, an LLM-orchestrated digital twin for distribution grid operations. An open-weights LLM orchestrates power system analysis and optimization solvers and, unlike prior systems, also performs uncertainty-aware studies: probabilistic security assessment, robust corrective dispatch, and flexibility-envelope and hosting-capacity characterization. We test it on the Bornholm 60 kV distribution network—a real Danish island power system—using one year of smart-meter measurements. An operator case study spans deterministic assessment, probabilistic risk quantification, and robust dispatch. Across a 68-prompt behavioral catalog scoring tool use, evidence consistency, state-mutation discipline, and refusal calibration, the orchestrator answers 98.5\% of tasks correctly on the first attempt—the lone failure being a missing answer, not a wrong one. The full pipeline is released open source.
\end{abstract}

\subsubsection*{Keywords:}
%Include up to five keywords that capture the main topics or themes of the paper. Separate each keyword with a comma and space.

power system operation, large language models, digital twin, uncertainty-aware analysis, distribution grid flexibility

\section{Introduction}

Generative artificial intelligence, and large language models (LLMs) in particular, has been adopted rapidly across domains such as healthcare, finance, manufacturing, and robotics, where it supports text and code generation, decision support, and complex problem solving \parencite{Alto2024LLMApps}. In parallel, the spread of real-time sensing in physical systems has accelerated the adoption of digital twins (DTs): virtual replicas continuously synchronized with measurement data that support monitoring, simulation, analysis, and optimization, including in power systems \parencite{Zhou2026DigitalTwinAI, Subramanian2026DigitalTwin}.

The key opportunity, and the focus of this work, is to extend such orchestration beyond deterministic single-point assessments toward \emph{uncertainty-aware} analyses driven by \emph{real operational measurements}---two capabilities that remain largely absent from existing LLM--DT frameworks. Operating a distribution grid increasingly requires not a single calculation but a sequence of analyses---state assessment, contingency screening, risk quantification under forecast uncertainty, and determination of corrective actions---each demanding the right analytical tool, correct parameterization, and proper interpretation of results. Driving these workflows manually is slow and demands significant domain expertise, with the complexity increasing further when probabilistic and robust analyses are considered. An LLM-orchestrated digital twin addresses this gap: the operator states an objective in natural language, and the LLM selects, configures, and chains \emph{validated} solvers and then interprets their outputs, preserving the human in the loop while reducing the burden of manually managing analytical workflows. Despite growing interest in LLM-integrated digital twins, their application to operational power systems remains relatively unexplored \parencite{Guo2025LLMDigitalTwinSurvey}.

Interest in applying LLMs to power systems has surged rapidly over the past two years \parencite{Zhou2026ComprehensiveReview}. Early work concentrated on information retrieval and result interpretation---querying system data and explaining simulation outcomes---without participating in control or optimization \parencite{Zhu2025PowerQA}. More recent work pushes the LLM toward decision-making and control: \textcite{Liu2025RePower}, for instance, uses an LLM directly as the optimizer for reactive-power compensation. Allowing the model itself to generate control decisions, however, raises safety and reliability concerns in high-risk operational environments, and such approaches remain confined to simulation studies. These concerns motivate the now-dominant paradigm, adopted here, of keeping humans in the loop and using the LLM as an assistant that orchestrates domain-specific tools rather than acting as an autonomous controller \parencite{Zhang2024OperationControl}. Because every result is produced by a validated solver whose invocation is explicitly reported by the agent, the path from request to answer remains auditable, preserving the transparency and human supervision required for operational deployment.

Within this assistant paradigm, several frameworks have appeared in quick succession. PowerDAG automates distribution-grid analyses by composing existing simulation tools \parencite{Badmus2026PowerDAG}; PFAgent couples intent parsing with AI-assisted evaluation to automate N-1 contingency analysis on IEEE benchmarks \parencite{She2026PFAgent}; GridMind integrates LLMs with deterministic solvers for AC optimal power flow and N-1 analysis \parencite{Jin2025GridMind}; X-GridAgent covers power flow, contingency, and optimal power flow analyses \parencite{Chen2025XGridAgent}; and Grid-Orch orchestrates a range of distribution-system studies using both local and cloud-hosted LLMs \parencite{Liu2026GridOrch}. Together these studies establish that LLMs can reliably drive conventional deterministic power-system computations through natural language interaction. Alongside these individual systems, the open-source PowerAgent community is assembling shared infrastructure for this paradigm---tool connectors, foundation models, and workflows \parencite{zhang2025poweragent}.
 
Despite this progress, the reviewed efforts above share two limitations. First, they are validated almost exclusively on synthetic or benchmark test systems rather than on real operational measurements. Second, and more fundamentally, their analytical capabilities are deterministic: they evaluate power flow, N-1 screening, and optimal power flow at a single assumed operating point, but do not support probabilistic security assessment, robust corrective optimization, or flexibility- and hosting-capacity characterization under uncertainty. Addressing these tasks requires more than repeated simulations. While thousands of deterministic power-flow evaluations can generate thousands of yes/no verdicts, they do not directly provide calibrated violation probabilities, risk-constrained corrective actions, or operational feasibility envelopes; each of those requires dedicated machinery---structured sampling, scenario-based robust optimization, and result aggregation---together with the orchestration logic to invoke and interpret it from a natural-language request. Because operational decisions are made under uncertainty in both load and renewable generation forecasts, the uncertainty-aware surface is essential for deployment yet remains absent from the reviewed orchestration frameworks.
 
This paper introduces CONDUCTOR, an LLM-orchestrated digital twin that addresses both gaps: it combines \emph{real} measurement-driven grid assessment with uncertainty-aware analysis and flexibility-oriented corrective action under operational constraints, and it is validated using real smart-meter measurements from an operational distribution network. Although the presented case study uses the Bornholm 60\,kV island distribution system, the analytical services are built on the open-source \code{pandapower} library \parencite{Thurner2018Pandapower}. As a result, the same orchestration layer and analytical tools can operate on any power system represented in \code{pandapower} format.
 
The contributions of this paper are threefold. (1)~\emph{Uncertainty-aware LLM orchestration}: a framework that drives operational power-system analyses from natural language---state assessment, N-1 contingency analysis, and flexibility-based optimization---and, distinctively, probabilistic security assessment, robust corrective optimization, and flexibility- and hosting-capacity characterization under uncertainty. (2)~\emph{Validation on real measurements}: an implementation on a measurement-driven digital twin that replays a full year of 15-minute smart-meter data as a stream of operating points, with every analysis executed on measured---rather than synthetic---grid states, through a network-agnostic tool layer. (3)~\emph{A fully open pipeline}: an open-source release of the orchestration layer, the tools, and demonstration case studies, built on an open-weights model (Gemma~4, Apache~2.0); because the weights are openly licensed, the orchestrator can be run either through the free hosted API used in this work or downloaded and served locally on the operator's own hardware without changing the whole pipeline.

\section{Framework}
\label{sec:framework}

\subsection{Architecture and orchestration principle}
\label{sec:architecture}

CONDUCTOR adds a natural-language orchestration layer on top of a measurement-driven digital twin of an operational distribution network. The underlying digital twin, established in our previous work (anonymized, et al., 2026), exposes three deterministic engines: the \emph{Real-Time Security Assessment Engine (\textbf{RSAE})}, which performs AC power-flow to assess voltage and thermal limits; the \emph{Contingency Analysis Engine (\textbf{CAE})}, which performs $N{-}1$ screening by simulating element outages and re-assessing system conditions; and the \emph{Security Management and Flexibility Activation Engine (\textbf{SMFAE})}, which provides corrective actions through optimization-based active and reactive power redispatch.

%CONDUCTOR adds a natural-language orchestration layer on top of a measurement-driven digital twin (DT) of an operational distribution network. The underlying DT, established in our previous work (anonymized, et al., 2026), exposes three deterministic engines:

%\emph{Real-Time Security Assessment Engine (RSAE)}: steady-state voltage and thermal screening.

%\emph{Contingency Analysis Engine (CAE)}: performs $N{-}1$ screening by simulating element outages and re-assessing security.

%\emph{Security Management and Flexibility Activation Engine (SMFAE)}: optimization-based corrective control through active and reactive-power redispatch.

These engines already provide advanced analysis, but interacting with them requires the user to manually select a function, configure engine-specific parameters, execute it, and interpret raw outputs---a workflow that scales poorly as the number of functionalities grows and that limits accessibility for non-expert operators.

%\iffalse
%\begin{figure}[t]
%    \centering
%	\includegraphics[trim={0.1cm 0.0cm 0.0cm 0.0cm}, clip,width=1\linewidth]{images/flow_update_1006.png}
%	\caption{Flow of the proposed LLM-orchestrated digital twin framework.}
%	\label{fig:framework}     
%\end{figure}
%\fi

\begin{figure}[t]
    \centering
	\includegraphics[trim={0.0cm 0.0cm 0.0cm 0.0cm}, clip,width=1\linewidth]{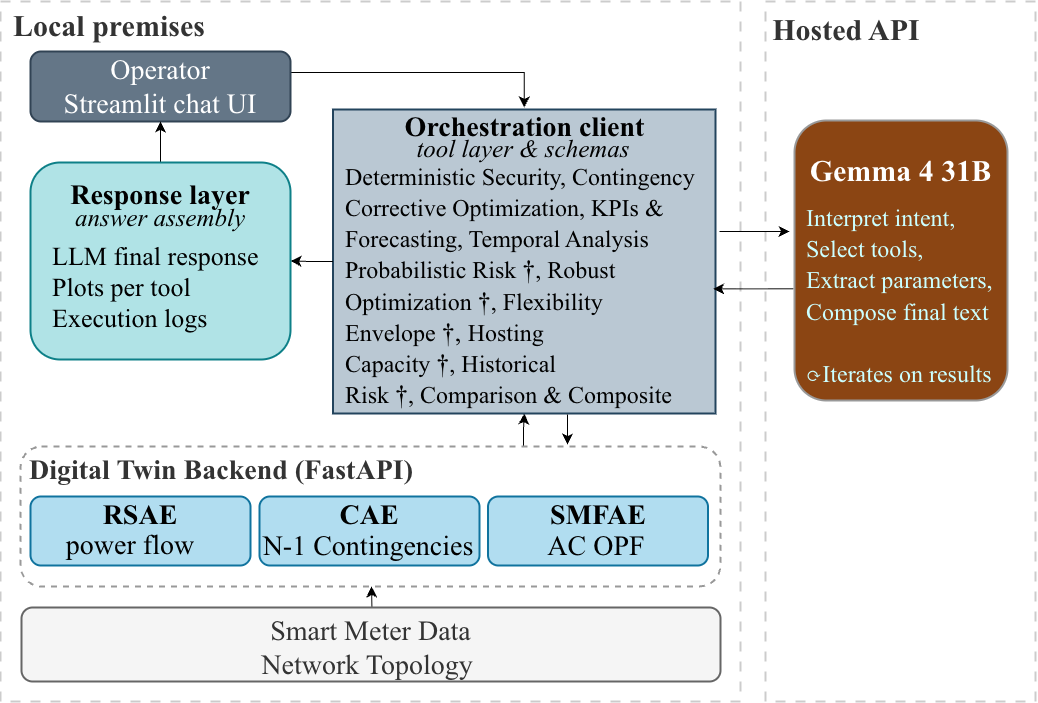}
	\caption{Flow of the proposed CONDUCTOR.}
	\label{fig:framework}     
\end{figure}

To remove this barrier, the numerical engines are left unchanged and instead treated as callable computational tools. A large language model (LLM) sits above them as an orchestration layer responsible for natural-language understanding, tool selection, parameter extraction, workflow coordination, and response generation, while all numerical results are produced by the corresponding digital twin engines using power system analysis and optimization solvers. Figure~\ref{fig:framework} summarizes the resulting workflow. The operator submits a natural-language query through the Streamlit chat interface, which is forwarded to the orchestrator. The \emph{Gemma 4 (31B)} model, an open-weights model developed by Google DeepMind under the Apache 2.0 license, interprets the operator's intent and maps it to one or more tools in the tool layer. The selected tools invoke the corresponding analytical engines in the digital twin backend, which execute the requested analyses using the current network topology and available measurement data. The resulting outputs are returned to the orchestration layer, where the LLM interprets and synthesizes the results into an operator-oriented response. Finally, the response layer formats and delivers the generated response with the plots and logs back to the operator. 

%Third-order headings, as in this paragraph, are discouraged in the template, but we allowed to use it of course.
%\paragraph{Data Privacy by Design.}

A central design principle of CONDUCTOR is the \emph{strict separation between orchestration and computation}. The LLM is limited to interpreting operator requests, selecting appropriate tools, and orchestrating their execution, while all engineering calculations are performed by validated deterministic digital twin engines. Consequently, the LLM never generates numerical results directly. This design ensures that all outputs remain physically grounded, reproducible, and fully traceable, reducing the risk of hallucinated quantities while preserving the reliability and interpretability required for operational use.

In this work, \emph{Gemma 4} is accessed through Google's hosted API, and all reported results use that configuration. By construction, the hosted model never receives the raw measurement database or the full network model: within the function-calling loop it sees only the operator's query, the tool schemas, and the structured results the tools return after executing locally---for example violation tables, dispatch records, or risk metrics. The complete smart meter time series, the \code{pandapower} model, and all numerical computations remain on the local backend. Thus, the information leaving the premises is limited to structured tool inputs and outputs rather than the operational data set itself. Because the weights are openly licensed, the model can be accessed through a free API tier, ensuring that the released tool remains reproducible for research purposes without requiring a paid subscription. Alternatively, the model can be deployed entirely on local hardware, requiring only modifications to the serving layer, thereby eliminating external data exchange. We note the latter as a supported deployment path rather than as a configuration evaluated here. The tool layer is model-agnostic, so the orchestrator can be replaced by any other function-calling LLM without modifying the engines.

\subsection{Tool layer and structured schemas}
\label{sec:tools}

Each digital twin functionality is exposed to the LLM as a structured tool with predefined input parameters, operational constraints, parameter descriptions, and a structured output. This constrains interaction to validated execution paths rather than free-form command generation. The tool schemas additionally encode the operational semantics of their parameters, including voltage-security thresholds, line and transformer loading limits, generator operating constraints, active and reactive external-grid import limits, generator-availability conditions, power-factor constraints, optimization weights, uncertainty parameters, and temporal horizons. This lets the LLM extract operational settings directly from a query and map them into a deterministic workflow. For example, a request for reduced cable import becomes a constraint on the allowable external-grid active-power exchange, and a generator-outage request triggers a topology modification
prior to analysis.

\begin{table*}[t]
\centering
\caption{Tool layer grouped by capability class. Classes marked $\dagger$
provide uncertainty-aware or empirical-historical analysis on the real
measurement stream.}
\label{tab:tools}
\footnotesize
\begin{tabular}{@{}lll@{}}
\toprule
Capability class & Representative tools & Backend engine / layer \\
\midrule
State \& deterministic security & \code{run\_rsa}, \code{get\_current\_conditions} & RSAE (power flow) \\
Contingency ($N{-}1$) & \code{simulate\_contingency}, \code{simulate\_all\_contingencies} & CAE \\
Corrective optimization & \code{optimize\_flexibility}, \code{optimize\_contingency} & SMFAE (AC OPF) \\
KPIs \& forecasting & \code{evaluate\_kpis}, \code{forecast\_kpis} & SMFAE (AC OPF) \\
Temporal analysis & \code{scan\_rsa\_over\_time}, \code{find\_worst\_case\_timestamp}, & RSAE over time \\
 & \code{get\_element\_timeseries}, \code{scan\_scenarios} & \\
Probabilistic risk $\dagger$ & \code{run\_probabilistic\_rsa} & RSAE + Monte Carlo (LHS) \\
Robust optimization $\dagger$ & \code{optimize\_robust\_flexibility} & Scenario / heuristic robust OPF \\
Flexibility envelope $\dagger$ & \code{compute\_flexibility\_envelope} & RSAE $(P,Q)$ sweep \\
Hosting capacity $\dagger$ & \code{compute\_hosting\_capacity} & RSAE bisection (det./prob.) \\
Historical risk $\dagger$ & \code{compute\_historical\_risk} & RSAE over smart meter window \\
Comparison \& composite & \code{compare\_results}, multi-tool workflows & Orchestration layer \\
\bottomrule
\end{tabular}
\end{table*}

Table~\ref{tab:tools} groups tools by capability class. Beyond the deterministic security, contingency, and corrective-optimization functions common to comparable toolkits, CONDUCTOR exposes five uncertainty-aware and empirical-historical functionalities ($\dagger$ in Table~\ref{tab:tools}), all operating on the same real measurement stream and forming the framework's principal analytical distinction. Most reuse the RSAE as their inner kernel---repeatedly evaluating the same security assessment across operating points or Monte Carlo samples and aggregating the result---so the uncertainty-aware surface is built on the validated deterministic engine rather than on new, unverified computation; robust optimization instead extends the AC-OPF.

\emph{Probabilistic security assessment} (\code{run\_probabilistic\_rsa}) propagates load and generation forecast uncertainty by Latin Hypercube sampling and reports per-bus violation probabilities with $P_5/P_{50}/P_{95}$ voltage envelopes. The sampled quantities are the uncertain forecast inputs---the demand that materializes and the renewable power actually available---not the operator's commanded setpoints, since forecast error perturbs what loads draw and what resources can deliver rather than the dispatch the operator chooses. These enter through the \code{load\_sigma} and \code{sgen\_sigma} parameters. \emph{Robust corrective optimization} (\code{optimize\_robust\_flexibility}) hardens a corrective dispatch against this uncertainty. A heuristic mode tightens per-bus voltage bounds from sampled percentile back-offs; a scenario mode solves a scenario-based AC-OPF whose sample count follows the scenario-approach bound of \textcite{calafiore2006scenario}, certified on an independent validation set. \emph{Flexibility envelope} (\code{compute\_flexibility\_envelope}) maps a generator's feasible $(P,Q)$ operating region. \emph{Hosting capacity} (\code{compute\_hosting\_capacity}) estimates incremental hosting capacity by bisection, in either a deterministic or a risk-capped probabilistic mode. \emph{Historical risk} (\code{compute\_historical\_risk}) characterizes empirical exceedance, violation episodes, and conditional risk over a smart-meter measurement window.

% We say this in several parts of the paper, we can reduce the lenght
%None of the LLM-integrated frameworks reviewed in Section~1 provides probabilistic, robust, or flexibility- and hosting-capacity analysis, and all are evaluated on synthetic or benchmark cases. The proposed framework delivers these capabilities on a real, continuously updated measurement stream.

\subsection{Natural-language routing and parameter extraction}
\label{sec:routing}

Routing over the three engines introduced in Section~\ref{sec:architecture} (RSAE, CAE, SMFAE) proceeds in four stages: (1) natural-language interpretation, (2) intent classification and tool selection, (3) parameter extraction, and (4) tool execution and response generation. Tool selection maps the query onto the DT engines---state assessment, voltage-security, and thermal-loading queries to the RSAE, contingency requests to the CAE, and corrective-action, flexibility, and optimization requests to the SMFAE. Stages (2)--(4) repeat within the function-calling loop: each structured result is returned to the LLM, which decides whether additional calls are required, enabling conditional branching (e.g., invoking robust optimization only when a probabilistic check fails), recovery from failed calls, and the decomposition of compound requests. The prompt \emph{``find the worst contingency during the next 24 hours and stabilize the network''}, for instance, expands into temporal scanning, contingency analysis, worst-case identification, flexibility optimization, and a corrective action summarization, all from a single interaction.

Within the same function-calling step, the LLM also populates each selected tool's arguments: guided by the parameter semantics declared in the tool schema, it extracts the relevant values from the query and maps them onto the engine inputs. The schema constrains this---argument types, allowed ranges, and required fields are enforced, so out-of-range or malformed values are rejected rather than executed, keeping interaction on validated paths. Four mappings recur: \emph{operating point}---the current timestamp is the default unless the operator specifies another; \emph{temporal horizon}---temporal queries set the analysis window; \emph{contingencies}---contingency requests identify the affected elements and outage scenarios; and \emph{topology}---generator-outage prompts trigger the required modification before analysis. Optimization requests additionally bind the SMFAE redispatch variables $P_g^{\text{base}}, Q_g^{\text{base}}$ and $P_g^{\text{new}}, Q_g^{\text{new}}$, the active and reactive outputs of generator $g$ before and after flexibility activation. 

In the final stage, the \emph{response layer} transforms the structured engine outputs---violation tables, redispatch records, and time-series---into operator-oriented summaries and plots. The natural-language response is generated by the same orchestrator LLM, without additional fine-tuning, while figures are produced by deterministic visualization routines associated with each tool. Thus, the plot follows from the tool invoked rather than from anything the model draws. The orchestration layer therefore remains a pure interface between natural-language interaction and deterministic computation, preserving the integrity of all the numerical results.

\subsection{Instantiation, portability, and availability}
\label{sec:instantiation}

The LLM orchestration layer and associated analytical services are designed to operate on any \code{pandapower}-compatible network model, supporting deployment across a wide range of grid topologies, from IEEE benchmark systems to real-world distribution networks. The underlying tool layer is network-agnostic and automatically transforms external network models into the required digital twin representation by constructing the necessary analytical structures and measurement interfaces. As a result, the same LLM-orchestrated workflow can be applied across different power systems without modifications to the user interaction layer or analytical services. The framework therefore combines operation on a real measurement-driven grid with portability across diverse network topologies.

To support transparency and reproducibility, CONDUCTOR is released as an open-source project\footnote{\url{https://github.com/antonioalcantaramata/CONDUCTOR}}. The release includes the deterministic engines, the structured tool layer and schemas, the LLM orchestration loop, the conversational interface, and a ready-to-run network and dataset, allowing the full pipeline to be reproduced and extended on openly available networks. Since Bornholm measurements consist of operational smart meter data, they cannot be redistributed. The results in this paper are therefore produced on the (non-redistributable) Bornholm digital twin, while the public release runs the identical code paths on a different, publicly available network with measurements and topology in \code{pandapower} format---keeping the entire stack, from analytical engines to language model, freely inspectable and re-runnable by other researchers and operators.

\section{Case Study on the Bornholm Distribution System}

CONDUCTOR is evaluated on the 60\,kV distribution network of Bornholm, a Danish island in the Baltic Sea. The island is supplied through a single sea-cable interconnection to the Swedish grid, which acts as the slack, and hosts substantial local wind generation connected at the 10\,kV level, so the network regularly transitions between import and reverse-flow export conditions. The digital twin is driven by real smart meter measurements obtained at 15-minute resolution, covering the full year 2022; the digital twin replays this record as a stream of operating points, and the conversation below begins at the first measurement of the year.
 
\begin{figure}[ht]
\centering
\includegraphics[width=\columnwidth]{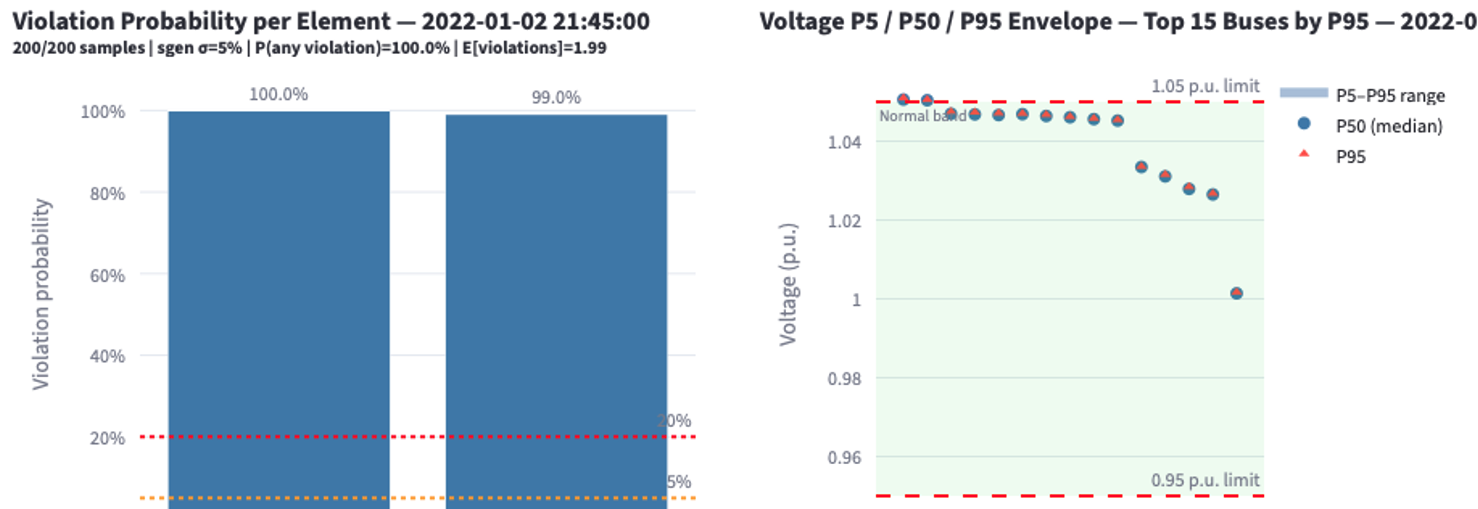}
\caption{Violation probability and voltage envelopes under $\pm$5\,\% load and generation uncertainty at the worst-case operating point (Act~3).}
\label{fig:prob-envelopes}
\end{figure}

The evaluation follows a single continuous conversation between a system operator and the agent, organized as six acts comprising eight natural-language prompts. The acts trace a realistic operational workflow: establish situational awareness, stress the security margins, look ahead in time, quantify confidence under forecast uncertainty, compute and verify a corrective action, harden it against larger uncertainty, and finally characterize local flexibility and hosting capability. Acts~3, 5, and~6 exercise precisely the uncertainty-aware analytical surface (probabilistic security assessment, robust corrective optimization, and flexibility-envelope and hosting-capacity characterization) that is absent from the LLM-integrated frameworks reviewed in Section~1. Because the digital twin replays a recorded year, the look-ahead in Act~2 operates on realized measurements under perfect foresight rather than on forecasts: it identifies the worst realized operating point in the week following the current timestamp. The forecast uncertainty an operator faces in real time is instead addressed explicitly in Acts~3 and~5, where estimation error enters through the probabilistic and robust tools; in deployment, the same search would run over forecast scenarios in place of the recorded stream, which the tools accept unchanged. Table~\ref{tab:workflow_rev} summarizes, for each prompt, the operator intent, the tool chain selected by the agent, and the reasoning the agent contributed beyond the explicit request. Throughout, the agent receives only the prompt: no tool is ever named by the operator. Selected agent responses are reproduced verbatim below, condensed by deletion only (marked [\ldots]); in addition to these textual summaries, the response layer renders plots systematically---each analytical tool is bound to a dedicated visualization routine, so the figures accompanying an answer are determined by the tools the LLM invokes rather than composed manually (Figure~\ref{fig:prob-envelopes} shows the visualization generated in Act~3).

\begin{table}[t]
\centering
\caption{Operator intent, the tool chain selected by the agent, and
the additional reasoning performed without explicit instruction.}
\label{tab:workflow_rev}
\small
\setlength{\tabcolsep}{3pt}
\renewcommand{\arraystretch}{1.15}
\begin{tabular}{@{}
  >{\raggedright\arraybackslash}p{0.18\columnwidth}
  >{\raggedright\arraybackslash}p{0.25\columnwidth}
  >{\raggedright\arraybackslash}p{0.45\columnwidth}@{}}
\toprule
\textbf{Intent} & \textbf{Tool chain} & \textbf{Added reasoning} \\
\midrule
Current operating condition &
$\cdot$ Timestamp $\cdot$ Conditions $\cdot$ RSA &
Inferred that situational awareness requires live measurements plus a
security assessment, with no explicit analysis request. \\
\addlinespace[3pt]
Security under tighter limits &
$\cdot$ RSA $\cdot$ N-1 screening &
Decomposed the compound request into N-0 and N-1, propagated the
custom voltage limit, and flagged the base case as already
infeasible. \\
\addlinespace[3pt]
Worst timestamp next week &
$\cdot$ Worst-case search $\cdot$ RSA $\cdot$ Conditions &
Translated ``next week'' into the forecast horizon and retrieved
context to explain the detected violations. \\
\addlinespace[3pt]
Confidence under 5\,\% uncertainty &
$\cdot$ Probabilistic RSA &
Mapped forecast error to probabilistic parameters and replaced
deterministic with probabilistic assessment. \\
\addlinespace[3pt]
Resolve infeasibility &
$\cdot$ RSA $\cdot$ Flexibility OPF $\cdot$ Comparison &
Verified the initial state, computed corrective actions, and
validated the solution with a before/after comparison. \\
\addlinespace[3pt]
Robustness under 10\,\% uncertainty &
$\cdot$ Probabilistic RSA $\cdot$ Robust OPF &
Conditional branching: evaluated feasibility first and invoked robust
optimization only upon failure. \\
\addlinespace[3pt]
Flexibility envelope of S2 &
$\cdot$ Timestamp $\cdot$ Envelope &
Resolved the temporal reference and identified the binding
operational constraint. \\
\addlinespace[3pt]
Hosting capacity of S2 &
$\cdot$ Hosting capacity $\cdot$ Retry &
Self-corrected an ambiguous bus name and turned a binary query into a
quantitative multi-strategy assessment. \\
\bottomrule
\end{tabular}
\end{table}

\medskip
\noindent\textbf{Act 1 --- Situational awareness.}
The shift begins with an open-ended status request:
 
\begin{operatorprompt}
\small\itshape ``Give me a summary of the current operating condition. Also include voltage margins and loading levels.''
\end{operatorprompt}
 
\noindent
The agent infers that situational awareness requires live data and a security assessment, and sequences \code{get\_current\_timestamp} $\rightarrow$ \code{get\_current\_conditions} $\rightarrow$ \code{run\_rsa} without being told to. At the current measurement (2022-01-01 00:00) the grid is secure: total load 6.04\,MW, local generation 4.05\,MW, 2.02\,MW imported from Sweden; all voltages within 0.95--1.05\,p.u.\ (range 0.998--1.049\,p.u.), maximum line loading 6.2\,\% (S1--S2~A) and maximum transformer loading 23.7\,\%. The operator then tightens the question:

\begin{operatorprompt}
\small\itshape ``Would the system still be secure if we want a maximum voltage of 1.045 and contingencies happen?''
\end{operatorprompt}
 
\noindent
The agent decomposes this compound request into an N-0 reassessment (\code{run\_rsa}) and a full N-1 screening (\code{simulate\_all\_contingencies}), propagating the custom limit \code{vm\_upper\_pu}\,$=$\,1.045 to both. Under the tighter ceiling the base case already exhibits eight overvoltage violations at 10\,kV buses (e.g., S2 10\,kV at 1.049\,p.u.), and all 40 simulated outages produce violations. Crucially, the agent concludes unprompted that the N-1 result is moot---the base case is already
infeasible, so corrective redispatch must precede any contingency security claim---an insight not stated in either tool output.

\medskip
\noindent\textbf{Act 2 --- Looking ahead.}
\begin{operatorprompt}
\small\itshape ``Consider the next week, find the worst timestamp and explain why the violations get worse.''
\end{operatorprompt}

\noindent
The agent translates ``next week'' into a 672-step horizon at the native 15-minute resolution, samples it via \code{find\_worst\_case\_timestamp} with violations as the search metric, navigates the twin to the result with \code{advance\_timestamp}, and then---because a violation count alone cannot answer ``why''---fetches both the security assessment and the operating conditions at that point. Its condensed answer:
 
\begin{agentresponse}
\small
The worst timestamp for grid security over the next week is 2022-01-02 21:45:00, which recorded 2 security violations [\ldots] S3 10.5\,kV: 1.051\,p.u.; S2 10kV: 1.050\,p.u.\ [\ldots] The two most active generators are located exactly where the violations occur. S3 is producing 3.54\,MW and S2 is producing 2.90\,MW [\ldots] the total system load is very low (6.31\,MW) [\ldots] the Sverige Slack is currently exporting power to the external grid (import of $-1.31$\,MW), which further reduces the amount of power being drawn from the local buses, allowing voltages to drift upward [\ldots] the grid becomes most stressed when renewable production peaks while demand is at a minimum, leading to voltage swells at the points of injection.
\end{agentresponse}
 
\noindent
The causal explanation---a low-load, high-injection export regime---is synthesized by the agent from two separate tool outputs; neither tool states it.

\medskip
\noindent\textbf{Act 3 --- Confidence under uncertainty.}
\begin{operatorprompt}
\small\itshape ``How confident are we that the system is secure or insecure at that specific timestamp, and which bus has the highest violation probability? Could you consider that we have 5\% estimation error in both generation and load?''
\end{operatorprompt}
 
\noindent
\noindent
A confidence question cannot be answered by the deterministic \code{run\_rsa} the agent has used so far; it switches to \code{run\_probabilistic\_rsa}, mapping ``5\,\% estimation error'' to \code{load\_sigma}\,$=$\,\code{sgen\_sigma}\,$=$\,0.05 and resolving ``that specific timestamp'' from the conversation context:
 
\begin{agentresponse}
\small
Probability of any violation: 1.00 (100\,\%) [\ldots] Expected number of violations: 1.99 [\ldots] Highest violation probability: S2 10kV (100\,\% probability of violation), followed closely by S3 10.5\,kV (99\,\%) [\ldots] Even in the most favorable 5\,\% of cases (P5), the voltage is 1.050\,p.u., which is already at the limit [\ldots] the insecurity is not a result of a ``worst-case'' fluke but is a systemic characteristic of this operating point.
\end{agentresponse}
 
\noindent
The interpretation in the final sentence goes beyond the raw percentiles: because the P5 voltage at both buses already sits at the limit, the violation probability of 1.00, an empirical estimate over the Latin-Hypercube sample (i.e., every sampled scenario violates), reflects a structural property of the operating point rather than a tail-risk outcome. The accompanying P5/P50/P95 voltage envelopes, rendered automatically by the visualization routine bound to \code{run\_probabilistic\_rsa}, are shown in Figure~\ref{fig:prob-envelopes}. The plots are generated automatically by the response layer: each analytical tool is mapped to a dedicated visualization routine, allowing the LLM to trigger the appropriate figure through tool selection. As a result, visualization becomes part of the orchestration workflow rather than a manually composed post-processing step.

\medskip
\noindent\textbf{Act 4 --- Corrective action.}
\begin{operatorprompt}
\small\itshape ``Solve the infeasibility for the mentioned worst timestamp.''
\end{operatorprompt}
 
\noindent
The agent re-verifies the pre-fix state (\code{run\_rsa}), computes the redispatch (\code{optimize\_flexibility}), and then closes the loop on its own initiative with \code{compare\_results}, producing a structured before/after diff. The fix is almost purely reactive: $\Delta Q$ of $-0.105$\,MVAr at S3 and $-0.049$\,MVAr at S2, compensated by $+0.279$\,MVAr from the interconnection, with all active-power changes below 0.01\,MW -- confirming, as the agent notes, a voltage-regulation problem rather than a power-balance one. Post-OPF, all buses return within the 0.95--1.05\,p.u.\ envelope.

\medskip
\noindent\textbf{Act 5 --- Making the fix robust.}
\begin{operatorprompt}
\small\itshape ``Is that dispatch still feasible under 10\% load uncertainty? If not, make it robust.''
\end{operatorprompt}
 
\noindent
The agent branches conditionally: it first audits the existing dispatch with \code{run\_probabilistic\_rsa} (\code{load\_sigma}\,$=$\,0.1) and escalates to \code{optimize\_robust\_flexibility} only because the audit fails. Its condensed answer:
 
\begin{agentresponse}
\small
No, the previous dispatch was not feasible under 10\,\% load uncertainty: the probabilistic assessment showed a 98.5\,\% probability of violation [\ldots] A robust OPF (heuristic back-off) tightened the voltage ceilings at S3 (1.050\,$\rightarrow$\,1.0493\,p.u.) and S2 (1.050\,$\rightarrow$\,1.0497\,p.u.), reducing reactive injection to 0.288\,MVAr (from 0.399) and 0.328\,MVAr (from 0.364) respectively [\ldots] After the robust OPF, the probability of any violation is 5.0\,\%, meeting the target risk.
\end{agentresponse}
 
\noindent
The two-step assess-then-act structure was entirely self-directed, as was the translation of ``10\,\% load uncertainty'' into the sampling parameters; the 95\,\% confidence target, however, is the tool default rather than an agent inference.

\medskip
\noindent\textbf{Act 6 --- Local capability.}
The operator finally drills into a single substation:
 
\begin{operatorprompt}
\small\itshape ``What is the flexibility envelope of S2 right now?'' \quad ``Does the S2 substation have hosting capacity?''
\end{operatorprompt}
 
\noindent
For the envelope, the agent resolves ``right now'' via \code{get\_current\_timestamp} and sweeps the $(P,Q)$ region with \code{compute\_flexibility\_envelope}: at the current dispatch of 2.11\,MW\,/\,0.30\,MVAr, the safe reactive range is $-1.02$ to $+0.34$\,MVAr, with 86 of 121 sampled points secure. The agent identifies the binding constraint unprompted: upper voltage limits, not thermal loading (maximum 32.1\,\%), cap the upward reactive flexibility. The hosting-capacity query exhibits self-correction: \code{compute\_hosting\_capacity} first fails on the ambiguous name ``S2'', and the agent retries with the qualified bus name ``S2 10kV'' without operator intervention. It also chooses \code{q\_mode}\,$=$\,\code{all} from the outset, turning a yes/no question into a quantitative comparison of reactive strategies: 1.71\,MW of hosting capacity at unity power factor (voltage-bound) versus 4.15\,MW under reactive-absorbing strategies (search-cap bound)---i.e., reactive control more than doubles the integration
headroom at this bus.

\medskip
Across the six acts, the agent decomposed compound requests, propagated user-defined constraints across successive analyses, resolved temporal and named-entity references from context, selected probabilistic over deterministic tools when the question demanded it, verified its own corrective actions, and recovered from a failed tool call---while every numerical quantity above was produced by the deterministic solvers of the digital twin, keeping all recommendations grounded in physically based, fully interpretable power system calculations. Unlike a conventional graphical interface, in which the operator must manually select each analysis, configure its parameters, and interpret raw outputs, the entire workflow was driven by high-level natural-language objectives and returned as actionable, explainable summaries with automatically generated plots. Acts~3, 5, and~6 in particular demonstrate, on measured operating points of a real network, the probabilistic, robust, and capability-characterization workflow that none of the LLM-integrated frameworks reviewed in
Section~1 provides.

\section{Behavioral Evaluation}
\label{sec:behavioral_evaluation}

Beyond the narrative case study, we evaluate CONDUCTOR on a catalog of 68 natural-language prompts spanning 15 task categories, from grid orientation and deterministic security assessment through probabilistic risk, robust optimization, and explicitly out-of-scope requests.

\subsection{Protocol}
\label{sec:eval-protocol}

Each catalog entry is a gradeable behavioral test rather than a bare query: it specifies the expected tool trajectory, explicit pass criteria, anticipated failure modes, a difficulty rating (14 easy, 38 medium, 16 hard), and a state-mutation flag indicating whether the prompt may legitimately move the simulation cursor. The 68 prompts are organized into 14 multi-turn conversations with 18 explicit history resets.

A turn is scored as a pass only when all of the following hold: (i) the invoked tool trajectory matches the expected tool(s); (ii) the final answer is consistent with the evidence returned by those tools; (iii) the agent's side effects match the state-mutation flag---a prompt marked non-mutating fails if the agent advances the operating point or launches heavier analyses than requested, even when the reported numbers are correct; and (iv) out-of-scope prompts are declined without fabricated capability. Criterion (iii) targets \emph{over-acting}, a failure mode of tool-using agents that standard answer-accuracy metrics do not capture. We report pass@1: each prompt is attempted once, with no retries, and scored on that single attempt. The orchestrator runs with greedy decoding (temperature~0), which is in principle deterministic; in practice, hosted inference on shared hardware can still vary slightly between runs. To characterize this, we executed the full catalog three times and report run-to-run consistency in Section~\ref{sec:eval-results}. Grading was performed manually by the authors against the predefined per-prompt criteria, with the expected tool trajectory checked against the execution logs.

% The orchestrator runs with greedy decoding (temperature~0), meaning it always selects the single most probable next token rather than sampling, which in principle makes its behavior deterministic. In practice, the model is served through a hosted API, and such cloud inference is not guaranteed to reproduce identical outputs from one run to the next---small numerical differences in how the computation executes on shared hardware can occasionally change the generated text. To characterize this, we executed the full catalog three times and report the run-to-run consistency in Section~\ref{sec:eval-results}.

\subsection{Results}
\label{sec:eval-results}

Table~\ref{tab:eval-results} reports per-category outcomes, which are identical across all three runs. In each run the agent passed 67 of 68 prompts (98.5\%), with 14 of the 15 categories at 100\%. Stratified by difficulty, pass@1 was 14/14 on easy, 38/38 on medium, and 15/16 on hard prompts in every run, the latter including multi-step compositions chaining up to four tools with intermediate-result reuse.

\begin{table}[t]
\centering
\caption{Per-category pass@1 on the 68-prompt behavioral catalog (identical in all three runs).}
\label{tab:eval-results}
\footnotesize
\setlength{\tabcolsep}{4pt}
\begin{tabular}{@{}lrrr@{}}
\toprule
Prompt category & Pass & Total & Pass\% \\
\midrule
A. Orientation \& Snapshot            & 3  & 3  & 100.0 \\
B. Deterministic Security (RSA)       & 7  & 7  & 100.0 \\
C. Contingency / N-1                  & 6  & 6  & 100.0 \\
D. Corrective Flexibility (OPF)       & 7  & 7  & 100.0 \\
E. Topology \& Generator Availability & 3  & 3  & 100.0 \\
F. Temporal Scans                     & 3  & 3  & 100.0 \\
G. Scenario Analysis                  & 2  & 2  & 100.0 \\
H. Probabilistic Risk                 & 4  & 4  & 100.0 \\
I. Robust Optimization                & 5  & 5  & 100.0 \\
J. Flexibility Envelope               & 3  & 3  & 100.0 \\
K. Hosting Capacity                   & 3  & 4  &  75.0 \\
L. Historical Risk                    & 4  & 4  & 100.0 \\
M. KPIs \& Forecasting                & 5  & 5  & 100.0 \\
N. Composite / Multi-Step             & 7  & 7  & 100.0 \\
P. Out of Scope / Not to Answer       & 5  & 5  & 100.0 \\
\midrule
\textbf{Total}                        & \textbf{67} & \textbf{68} & \textbf{98.5} \\
\bottomrule
\end{tabular}
\end{table}

We assess consistency by comparing, for each of the 68 prompts, its three repeated executions. Stability is high but not exact: every prompt received the same pass/fail verdict in all three runs, and 49 of the 68 produced word-for-word identical answers all three times. The remaining variation was minor---one prompt varied in its number of tool calls and two in their tool path---and never breached state-mutation discipline: prompts flagged non-mutating never moved the simulation cursor. It arose only on prompts permitted to advance the cursor, where repeated runs can legitimately settle on slightly different but equally valid states; in every case the answer remained consistent with the tool evidence of its own run.

Supported tasks were generally completed with the expected tools and evidence-consistent answers, including follow-ups that reused prior results when conversation context was preserved. The single hard failure fell in the hosting-capacity category (K, 3/4) and recurred identically in all three runs: the turn produced no tool activity and no final answer, the model stalling during its reasoning step. Notably, this was a fail-silent rather than fail-wrong outcome---across the three runs (204 prompt executions), the agent never returned an answer inconsistent with its tool evidence, the only observed failure mode being the absence of an answer. For an operator-facing decision support system, this asymmetry matters: a missing answer is recoverable by a retry, whereas a confidently wrong number is not.

\subsection{Effect of the System Prompt}
\label{subsec:system-prompt}
 
The agent's \emph{system prompt} is the standing instruction block defining its role and operating discipline: how to select tools, when to decline out-of-scope requests, the requirement to disclose which tool it called, and the injunction not to act beyond what was asked. To isolate what this block contributes, we re-executed the full catalog three times with it removed, leaving the model only the tool schemas (the machine-readable descriptions of each tool and its arguments). Everything else---the 68 prompts, the tools, the decoding settings---was held fixed.

Overall accuracy is unchanged: the ablated agent also passes 67 of 68 prompts in every run, expected because the tool descriptions alone carry enough information to solve most tasks. What changes is \emph{which} prompt fails. With the system prompt, the single failure is the hosting-capacity stall described above. Without it, the agent failed on a different prompt: an intentionally unsupported request (generator unavailability combined with N-1 screening) that it should decline. Lacking the instruction to refuse, it improvises a workaround and reports a result as if the combined analysis were valid. This swap is reproducible---the same prompt fails silently with the system prompt, and the same different prompt fails with a wrong answer without it, in all three matched run pairs. Two conclusions follow. First, the system prompt does not raise the pass rate on supported tasks (67/68 either way). Second, and more important, it governs \emph{how the agent fails at the edge of its capabilities}: with the prompt, an unsatisfiable request yields no answer (fail-silent); without it, the same class of request yields a confident but wrong one (fail-wrong). The system prompt therefore shifts the boundary failure mode onto the recoverable side of the asymmetry noted above---a missing answer can be caught and retried, a fabricated one may not be.

The system prompt's remaining contribution is transparency and procedural discipline. (The agent still functions without it---it receives the user's request and the tool schemas, which suffice to operate.) With the system prompt, the agent disclosed which tool it called and with what parameters in 178 of 204 responses, versus 0 of 204 without it, making the path from computation to answer auditable from the response alone. It committed 3 tool errors across all runs versus 12 without the system prompt, and was more stable under repetition (49 vs.\ 43 of 68 word-for-word identical responses; 2 vs.\ 4 tool-path changes). In exchange, the ablated agent is faster and leaner (mean 43\,s vs. 68\,s per prompt, fewer total tool calls)---a more opportunistic mode, adequate on mainstream supported tasks but undisciplined at capability boundaries.

\paragraph{Limitations.} Some limitations should be considered when interpreting the results. The behavioral evaluation catalog was developed by the authors, and an independent benchmark would provide stronger external validity. Validation was conducted on the Bornholm distribution network; although the tool layer is network-agnostic and supports arbitrary \code{pandapower}-based systems, evaluation on additional real-world networks remains future work. The study also considers a single orchestrator model, and alternative LLMs may exhibit different performance and failure modes. Finally, behavioral scoring relied on manual assessment supported by execution-log verification, while long-term deployment aspects such as operator trust, user adoption, and organizational integration remain future work.

\section{Conclusion}
\label{sec:concl}

This paper presented CONDUCTOR, an LLM-orchestrated digital twin that places a natural-language orchestration layer above validated engineering solvers for distribution grid operations. Its distinguishing contribution is twofold. First, CONDUCTOR exposes uncertainty-aware analyses that deterministic frameworks omit: probabilistic security assessment, robust corrective optimization, and flexibility-envelope and hosting-capacity characterization. Second, these analyses are validated on a real 60 kV island distribution network—Bornholm, Denmark—driven by a full year of operational smart-meter measurements, rather than on synthetic test cases. Through eight natural-language prompts, the case study followed an operator from situational awareness to a robustly secured dispatch, with the agent decomposing compound requests, propagating constraints across analyses, choosing probabilistic over deterministic tools when warranted, and verifying its own corrective actions. A separate 68-prompt behavioral catalog then evaluated the orchestrator: 67 passed on the first attempt (98.5\%), the agent neither over- nor under-refused, and its single failure was an absent answer rather than a wrong one.

This asymmetry is not incidental—it follows from a deliberate design choice. The LLM acts as an orchestrator---interpreting intent, routing to a tool, and reporting the result---and never computes engineering quantities, so every number it returns comes from a validated solver. Having no channel to invent a quantity, the model tends to fail by not answering rather than by returning a plausible but wrong value. For operator-facing systems, this distinction is critical: a stalled response is recoverable, whereas a confidently wrong number may go unchecked. Future work will expand open repository functionalities, including fully local model serving.

\section*{Acknowledgement}
\label{sec:acknow}

This work was supported by the Horizon-IA project ODEON, Grant Agreement 101136128, funded by HORIZON Actions within the European Union.

%\section{Appendices}

% if added before the last page, this command can help balancing columns
%\addtolength{\textheight}{-.2cm} 

%Bibliography 
% \bibliographystyle{apalike}
% \bibliography{sample}

\printbibliography

\end{document}